\documentclass[aip,prl,reprint]{revtex4-2}
\usepackage[T1]{fontenc}
\usepackage{times}
\usepackage{graphicx}
\usepackage{booktabs}

\usepackage{amsfonts, amsmath,amssymb, bm, graphicx}
\usepackage{siunitx}

\usepackage{amssymb}
\usepackage[english]{babel}
\usepackage{lipsum}

\newcommand{\figref}[2]{\hyperref[#1]{\ref{#1}(#2)}}
\newcommand{\figrefsub}[3]{\hyperref[#1]{\ref{#1}(#2)#3}}

\usepackage{physics}
\usepackage{lipsum}
\usepackage{changes}

\usepackage[colorlinks=true,allcolors=blue]{hyperref}
\usepackage{footmisc}

\usepackage{upgreek}

\usepackage{soul}

\makeatletter
\let\ORIbbl@fixname\bbl@fixname
\def\bbl@fixname#1{%
  \@ifundefined{languagealias@\expandafter\string#1}
    {\ORIbbl@fixname#1}
    {\edef\languagename{\@nameuse{languagealias@#1}}}%
}
\newcommand{\definelanguagealias}[2]{%
  \@namedef{languagealias@#1}{#2}%
}
\makeatother

\definelanguagealias{en}{english}

\begin{document}

\title{Modification of three-magnon splitting in a flexed magnetic vortex}

\author{L. K\"orber}\email{l.koerber@hzdr.de}
\affiliation{Helmholtz-Zentrum Dresden--Rossendorf, Institut f\"ur Ionenstrahlphysik und Materialforschung, D-01328 Dresden, Germany}
\affiliation{Fakultät Physik, Technische Universität Dresden, D-01062 Dresden, Germany}

\author{C. Heins}
\affiliation{Helmholtz-Zentrum Dresden--Rossendorf, Institut f\"ur Ionenstrahlphysik und Materialforschung, D-01328 Dresden, Germany}
\affiliation{Fakultät Physik, Technische Universität Dresden, D-01062 Dresden, Germany}

\author{I. Soldatov}
\affiliation{Institute for Integrative Nanosciences, Leibniz Institute for Solid State and Materials Science (IFW) Dresden, Helmholtzstrasse 20, 01069 Dresden, Germany}

\author{R. Schäfer}
\affiliation{Institute for Integrative Nanosciences, Leibniz Institute for Solid State and Materials Science (IFW) Dresden, Helmholtzstrasse 20, 01069 Dresden, Germany}
\affiliation{Institut für Materialphysik, Technische Universität Dresden, D-01062 Dresden, Germany}

\author{A. Kákay}
\affiliation{Helmholtz-Zentrum Dresden--Rossendorf, Institut f\"ur Ionenstrahlphysik und Materialforschung, D-01328 Dresden, Germany}

\author{H. Schultheiss}
\affiliation{Helmholtz-Zentrum Dresden--Rossendorf, Institut f\"ur Ionenstrahlphysik und Materialforschung, D-01328 Dresden, Germany}
\affiliation{Fakultät Physik, Technische Universität Dresden, D-01062 Dresden, Germany}

\author{K. Schultheiss}
\affiliation{Helmholtz-Zentrum Dresden--Rossendorf, Institut f\"ur Ionenstrahlphysik und Materialforschung, D-01328 Dresden, Germany}








\date{\today}

\begin{abstract}

We present an experimental and numerical study of three-magnon splitting in a micrometer-sized magnetic disk with the vortex state  strongly deformed by static in-plane magnetic fields. Excited with a large enough power at frequency $f_\mathrm{RF}$, the primary radial magnon modes of a cylindrical magnetic vortex can decay into secondary azimuthal modes via spontaneous three-magnon splitting. This nonlinear process exhibits selection rules leading to well-defined and distinct frequencies $f_\mathrm{RF}/2\pm \Delta f$ of the secondary modes.
Here, we demonstrate that three-magnon splitting in vortices can be significantly modified by deforming the magnetic vortex with in-plane magnetic fields, leading to a much richer three-magnon response. We find that, with increasing field, an additional class of secondary modes is excited which are localized to the highly-flexed regions adjacent to the displaced vortex core. While these modes satisfy the same selection rules of three-magnon splitting, they exhibit a much lower three-magnon threshold power compared to regular secondary modes of a centered vortex. The applied static magnetic fields are small ($\simeq \SI{10}{\milli\tesla}$), providing an effective parameter to control the nonlinear spectral response of confined vortices. Our work expands the understanding of nonlinear magnon dynamics in vortices and advertises these for potential neuromorphic applications based on magnons.

\end{abstract}

\maketitle

Magnetic vortices are non-uniform magnetization states which naturally appear in condensed matter and are of interest for both fundamental as well as applied research. In thin easy-plane ferromagnets, they are comprised by an outer region, the vortex \textit{skirt}, where the direction of magnetization $\bm{m}=\bm{M}/M_\mathrm{s}$ (with $M_\mathrm{s}$ being the saturation magnetization of the material) performs a full $2\pi$ rotation in the plane of the ferromagnet, and an inner region, the vortex \textit{core}, where the magnetization points out of the plane at the center of rotation. Topologically, vortices are merons, closely related to Skyrmions, with a half-integer topological charge.\cite{kosevichMagneticSolitons1990,metlovTwodimensionalTopologicalSolitons2001} Isolated vortices can be observed, for example, in nano-to-micrometer-sized flat magnetic disks. In such systems, vortices have been proposed for example as storage devices, encoding information into the vortex-core polarity (up or down).\cite{shinjoMagneticVortexCore2000,vanwaeyenbergeMagneticVortexCore2006,hertelUltrafastNanomagneticToggle2007} Other works propose vortex-resonators based on the gyrotropic motion of the core.\cite{gaidideiMagneticVortexDynamics2010,guslienkoEigenfrequenciesVortexState2002,pribiagMagneticVortexOscillator2007,mistralCurrentDrivenVortexOscillations2008,ruotoloPhaselockingMagneticVortices2009} Apart from the statics and dynamics related to the core, vortices also bear a rich spectrum of azimuthal and radial magnon modes, which live, predominantly, in the vortex skirt.\cite{ivanovMagnonModesMagnonvortex1998,shekaAmplitudesMagnonScattering2004,ivanovHighFrequencyModes2005,buessExcitationsNegativeDispersion2005} These modes exhibit interesting linear as well as nonlinear dynamics. For disk radii in the micrometer regime, the magnon spectrum allows for resonant three-magnon splitting (3MS), where one primary magnon, excited above a certain power threshold, splits into two secondary magnons under conservation of energy and angular momentum.\cite{lvovWaveTurbulenceParametric1994} Recently, it has been shown how 3MS can be exploited to excite modes with unprecedented large azimuthal mode number,\cite{schultheissExcitationWhisperingGallery} which can potentially couple to photon modes in optical cavities. In magnetic vortices, 3MS exhibits special selection rules, leading to secondary modes with distinct frequencies,\cite{verbaTheoryThreemagnonInteraction2021} and can even be stimulated non-locally below its nonlinear threshold.\cite{korberNonlocalStimulationThreemagnon2020} These qualities make confined vortices attractive for nonlinear networks and reservoir computing, for example, as a magnon-scattering reservoir.\cite{korberNonlocalStimulationThreemagnon2020,korberPatternRecognitionMagnonscattering2022}

So far, 3MS in confined vortices has been studied in the absence of external fields, where the vortex core is centered within the magnetic disk. In this work, we study the influence on the nonlinear magnon dynamics of an in-plane magnetic field which displaces the vortex core from its center position and leads to a flexing of the vortex skirt. Our study is carried out numerically using micromagnetic simulations and experimentally using Kerr microscopy and micro-focused Brillouin-light-scattering spectroscopy. We find that the spectrum of secondary modes acquires additional distinct frequency features, originating to two different kinds of magnons, localized either to the quasi-homogeneous (field-aligned) or the highly-flexed regions of the skirt. The latter ones, which only exist in a flexed vortex, are found to exhibit a significantly lower three-magnon threshold power compared to the regular modes in the field-free case.
Despite breaking the cylindrical symmetry of the magnetic state, the selection rules of 3MS in vortices persist, providing secondary modes with well-defined frequencies. The in-plane magnetic fields necessary to achieve a sufficient vortex deformation are small ($\simeq \SI{10}{\milli\tesla}$) and could be generated with on-chip stripline antennae.\cite{schultheissTimeRefractionSpin2021} This work contributes to the understanding of magnon modes in a magnetic vortex and explores an effective parameter to modify their nonlinear characteristics \textit{in-situ}.

\begin{figure}
    \centering
    \includegraphics{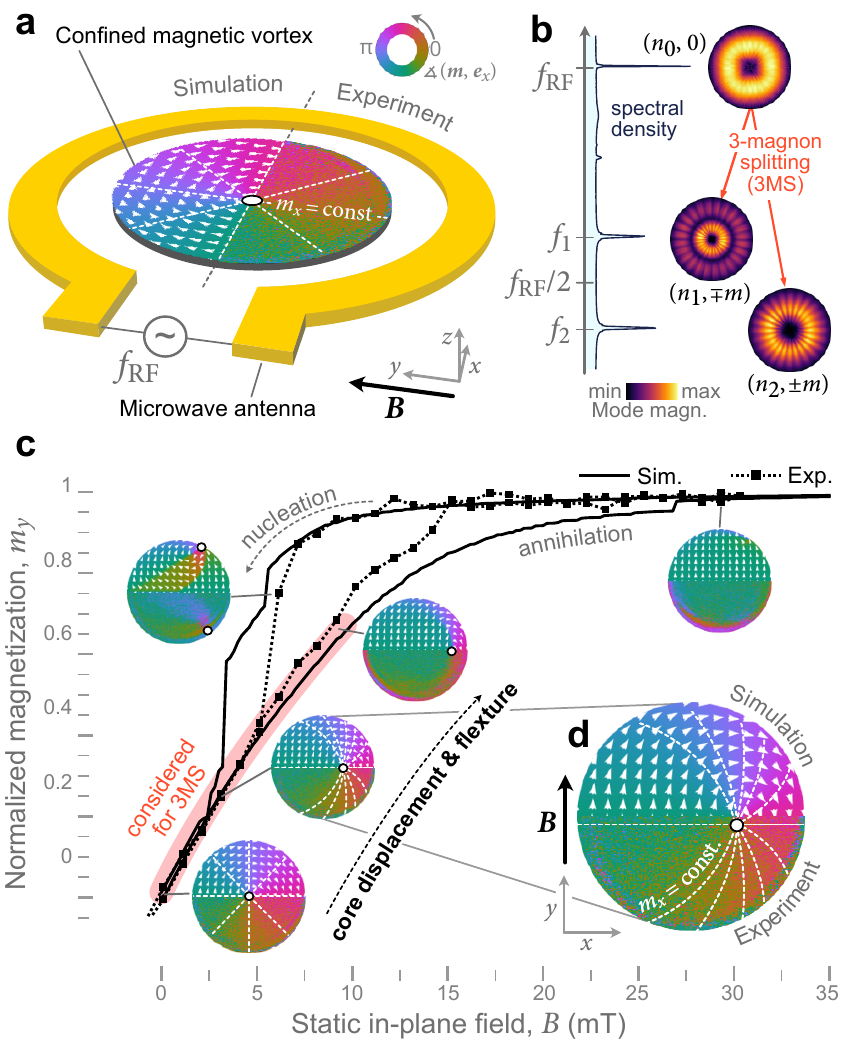}
     \caption{(a) Schematics of a magnetic Py disk with \SI{5.1}{\micro\meter} diameter and \SI{50}{\nano\meter} thickness and of the antenna design used in the experiments. At remanence, the disk exhibits a stable vortex state, confirmed with micromagnetic simulations and, experimentally, with Kerr microscopy. The color code depicts the in-plane angle of the magnetization, obtained with both methods. A white circle indicates the position of the vortex core where the magnetization tilts out-of-plane. Dashed lines denote the contour lines of constant $m_x$ ($x$ component of the normalized magnetization) obtained from the simulations. (b) Exciting a radial mode above the threshold triggers nonlinear three-magnon splitting, and pairs of modes with frequencies $f_{1,2}=f_\mathrm{RF}/2 \pm \Delta f$ and different radial profiles are  excited. The shown spatial mode profiles have been obtained from micromagnetic simulations. (c) Measured (dotted line) and simulated (solid line) hysteresis loop segment of the vortex disk with in-plane external field is shown. The magnetization distribution for certain field values is added as insets, in which the color code represents the experimentally and numerically obtained in-plane angle of the magnetization states. In particular, the lower half of each inset shows the respective quantitative Kerr image.}
    \label{fig:FIG1}
\end{figure}

For our study, we consider a magnetic disk of \SI{50}{\nano\meter} thick Ni$_{81}$Fe$_{19}$ (permalloy, Py) with a diameter of \SI{5.1}{\micro\meter},\cite{schultheissExcitationWhisperingGallery} as seen in Fig.~\figref{fig:FIG1}{a}. In such a system, a magnetic vortex is stable and corresponds to the state which minimizes magnetic strays fields everywhere except in the core region. The presence of the vortex state at zero field is confirmed with micromagnetic simulations\cite{schultheissExcitationWhisperingGallery,korberNonlocalStimulationThreemagnon2020,vansteenkisteDesignVerificationMuMax32014} and Kerr microscopy.\cite{hubertMagneticDomainsAnalysis1998,soldatovSelectiveSensitivityKerr2017} In Fig.~\figref{fig:FIG1}{a}, we present the in-plane angle of the magnetization obtained with both methods as a color map superimposed on the top surface of the magnetic disk.

To discuss the deformation of the vortex under application of an in-plane external magnetic field $\bm{B}$ (here, applied in $y$ direction), it is useful to consider the contour lines of the magnetization component $m_x=\mathrm{const}.$ perpendicular to the magnetic field. In the following, we only show the contour lines (as dotted lines) obtained from the numerical data. At zero field, these contours are straight lines, in agreement with an absence of magnetic volume charges $\varrho=-M_\mathrm{s}\bm{\nabla}\bm{m}$. With increasing vortex-displacement, these charges become nonzero and play an important role for the magnon dynamics. In field-free case, the magnon modes can be characterized by their number of nodal lines $n=0,1,2,...$ in radial direction and their number of periods $m=0,\pm 1,\pm 2$ in azimuthal direction. An $\Omega$-shaped microwave antenna is used to excite radial magnon modes with $m=0$ at frequency $f_\mathrm{RF}$. When the input power is high enough, this primary radial mode splits into two secondary modes with opposite azimuthal indices $\pm m$ (conservation of angular momentum), distributed around half the excitation frequency $f_{1,2}=f_\mathrm{RF}/2\pm \Delta f$ (conservation of energy), as seen in Fig.~\figref{fig:FIG1}{b}. An additional selection rule requires the two secondary modes to exhibit different radial indices, resulting in a frequency split $\Delta f \neq 0$ between them, as shown by~\textcite{schultheissExcitationWhisperingGallery} \textcolor{black}{In first order of perturbation (with respect to excitation power), the split $\Delta f$ is determined by the band gap at given azimuthal index $m$ between modes with different radial index $n$. For larger excitation powers, this split is altered by nonlinear frequency shift.\cite{krivosikHamiltonianFormulationNonlinear2010}}

Before we explore the influence of vortex deformation on 3MS, we study the effect of an externally applied magnetic field on the vortex state itself, in particular, its in-plane hysteresis loop. For this, we sweep the external field from 0 to \SI{40}{\milli\tesla} (and back), tracking the vortex state with micromagnetic simulations and Kerr microscopy. In Fig.~\figref{fig:FIG1}{c}, we show the evolution of the magnetization component $m_y$ parallel to the applied field together with two-dimensional maps of the vortex state at certain points of the loop. For moderate in-plane fields (below \SI{15}{\milli\tesla}), the vortex core is displaced from its center position which leads to a growth of the region of the vortex skirt magnetized parallel to the field (green area in the angle maps). However, in parallel to this displacement of the core, the skirt of the vortex is not simply displaced transversally to the field (which would correspond to a rigid-vortex model). Instead, the vortex skirt flexes in order to keep the magnetization parallel to the sides of the disk which avoids magnetic edge charges and, thus, minimizes stray fields. This can be seen nicely in Fig.~\figref{fig:FIG1}{d}, as the contour lines of $m_x$ clearly bend (deviating from rigid core displacement). Such behavior is described by the two-vortex model\cite{guslienkoEvolutionStabilityMagnetic2001,metlovStabilityMagneticVortex2002} for zero field, or, more appropriate for the case of applied fields, by a deformable-vortex pinning model.\cite{burgessAnalyticalModelVortex2014} A description of the vortex displacement under in-plane magnetic fields in similar magnetic elements is found, \textit{e.g.} in Refs.~\citenum{schaeferHysteresisSoftFerromagnetic2002,desimoneLowEnergyDomain2002}.

Finally, after further increasing the applied magnetic field, the vortex core reaches the boundary of the disk and annihilates, allowing for a saturation of the magnetic sample. When again decreasing the external field from saturation, two vortices nucleate at the boundary, which move together and finally merge to a single vortex close to zero applied field. We highlight that, in Fig.~\figref{fig:FIG1}{d}, this two-vortex nucleation is seen in both experiment and simulation, attesting to the good agreement between both methods.

\begin{figure}
    \centering
    \includegraphics{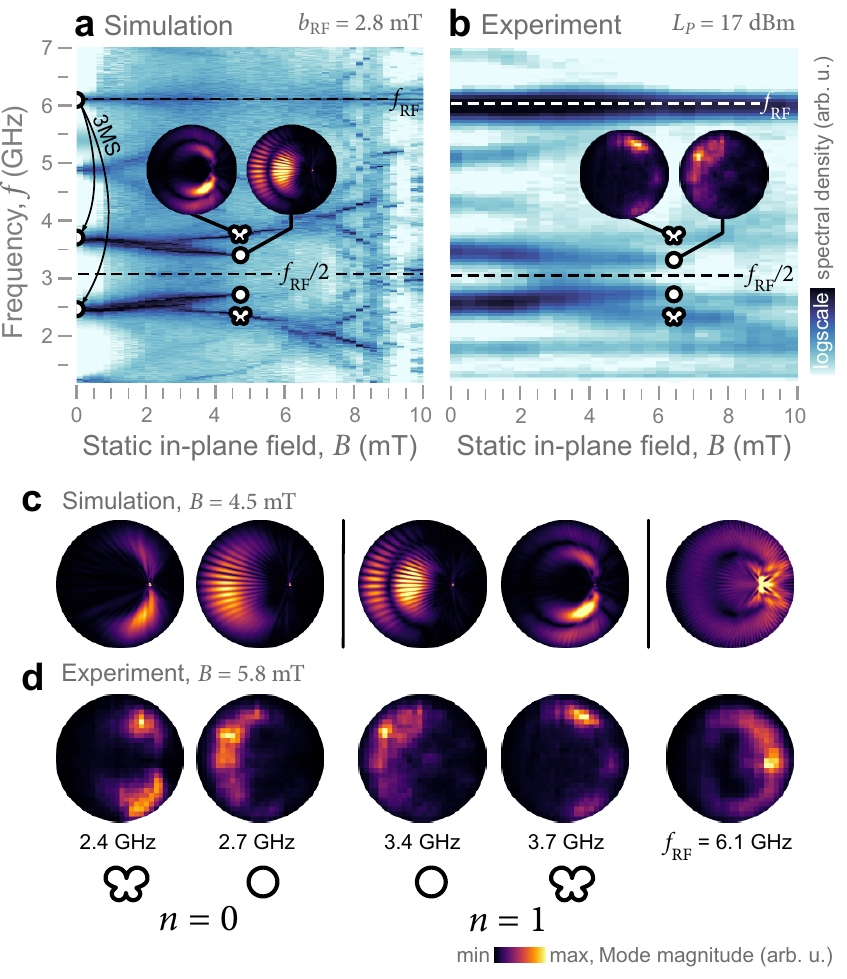}
    \caption{(a) Numerically and (b) experimentally obtained frequency response of the magnetic vortex excited at $f_\mathrm{RF}=\SI{6.1}{\giga\hertz}$ above the threshold for 3MS as a function of the applied in-plane magnetic field. At zero field, only one pair of secondary modes is created around $f_\mathrm{RF}/2$ by 3MS (other signals at zero field are only higher-harmonics\cite{schultheissExcitationWhisperingGallery}). With increasing in-plane magnetic field, the vortex is deformed which leads to a splitting of the three-magnon response into several branches, corresponding to the \textit{regular} vortex modes and additional \textit{butterfly} modes, as seen in the mode profiles. (c) Numerical and (d) experimental profiles of all modes at a given vortex-displacement (in-plane field), showing that regular and butterfly modes are arranged in pairs which satisfy the selection rules of 3MS in vortices.}
    \label{fig:FIG2}
\end{figure}

For the following study of nonlinear magnon interaction, we shall restrict ourselves to the field range from $0\rightarrow\SI{10}{\milli\tesla}$ in which the vortex core is merely displaced, as marked in Fig.~\figref{fig:FIG1}{c}. We start again at zero field and excite the confined vortex with an out-of-plane microwave field at $f_\mathrm{RF}=\SI{6.1}{\giga\hertz}$. In the micromagnetic simulations, the strength of the field is set to $b_\mathrm{RF}=\SI{2.8}{\milli\tesla}$. In our experiments, we set the microwave output power to $L_P = 17$\,dBm. Even though it is experimentally cumbersome to determine the exact microwave power arriving at the sample and, thus, comparing it with the field applied in the simulations, in both cases, the radial mode $(n=0, m=0)$ is excited above its 3MS threshold. In experiments, the spectral response of the system is probed by means of micro-focused Brillouin-light-scattering spectroscopy\cite{sebastianMicrofocusedBrillouinLight2015} and in simulations, by means of Fourier analysis. As seen in Fig.~\figref{fig:FIG2}{a}, nonlinear splitting of the excited primary mode leads to two secondary modes around half the excitation frequency. Additional frequency contributions can be attributed to higher-order processes (\textit{e.g.} four-magnon-scattering between the secondary modes) and, for simplicity, will not be considered in the following discussion.

To investigate the field dependence of the spectral response, we increase the external field to displace the vortex core from its center position, while maintaining the microwave excitation at \SI{6.1}{\giga\hertz}. The corresponding spectral response as a function of applied field is shown as colormaps in Fig.~\figref{fig:FIG2}{a} (simulation) and Fig.~\figref{fig:FIG2}{b} (experiment). Discrepancies between simulation and experiment can be attributed to the previously discussed uncertainty in the microwave-power levels as well as a small uncertainty ($< \SI{1}{\milli\tesla}$) in the experimentally set in-plane fields. For very small fields ($\lesssim \SI{2.5}{\milli\tesla}$) we see that the frequency split $\Delta f$ between the secondary modes decreases slightly. As a first observation, overall, 3MS is found to be robust against vortex deformation. However, with increasing in-plane field [see again Figs.~\figref{fig:FIG2}{a,b}], we observe a splitting of the secondary modes into several branches. This splitting continues for even higher in-plane fields. To identify the different branches, we obtain the spatial modes profiles at $B=\SI{4.5}{\milli\tesla}$ in the simulations by means of inverse Fourier transform at the respective frequencies. In our experiments, the spatial mode profiles are obtained at $B=\SI{5.8}{\milli\tesla}$ by scanning the whole sample with BLS microscopy.\cite{sebastianMicrofocusedBrillouinLight2015} \textcolor{black}{These two external fields were chosen to achieve well-separated peaks in the frequency spectra obtained with each method. Their difference can again be justified by the uncertainty in microwave power and external field mentioned above, as well as by the lower spectral resolution in the experiments.} \textcolor{black}{The numerical mode magnitude corresponds to the local magnitude $\vert \Tilde{m}_z(\bm{r},f)\vert$ of the corresponding Fourier component of the dynamic magnetization, which is proportional to the photon counts we measure in our BLS experiments.}

The profiles of all modes (directly excited and secondary) are shown for both methods in Figs.~\figref{fig:FIG2}{c,d}. We find that the different branches correspond to two qualitatively distinct classes of modes. The first class of modes resembles the \textit{regular} modes of a magnetic vortex and are merely deformed versions of the modes in a centered vortex~\cite{schultheissExcitationWhisperingGallery}. For the chosen magnetic field, these modes oscillate at $\SI{2.7}{\giga\hertz}$ and at $\SI{3.4}{\giga\hertz}$. The second class of modes is characterized by unconventional spatial profiles, resembling the shape of a butterfly with the vortex core at its center ($\SI{2.4}{\giga\hertz}$ and $\SI{3.7}{\giga\hertz}$). Clearly, these modes are arranged in separate pairs, a pair of regular secondary modes and a pair of butterfly secondary modes. Each of these pairs satisfies the three-magnon resonance condition $f_{1,2}=f_\mathrm{RF}/2\pm \Delta f$. It is quite remarkable that, even-though the cylindrical symmetry of the system is broken and a characterization of the modes in terms of radial and azimuthal mode numbers becomes ambiguous, the secondary modes within one scattering channel (regular or butterfly) still obey to certain selection rules which lead to a non-zero frequency split between them. In the numerically obtained mode profiles in Fig.~\figref{fig:FIG2}{c}, it can be nicely seen how the modes in each pair still have a different number of nodal lines along the new "radial" direction. This retaining asymmetry between the spatial profiles of the secondary modes is not surprising, as the vortex still inherits a mirror symmetry even when deformed into a flexed state. Such mirror symmetry alone can lead to selection rules for three-magnon splitting, as it is the case in magnetic films.\cite{lvovWaveTurbulenceParametric1994}

\begin{figure}
    \centering
    \includegraphics{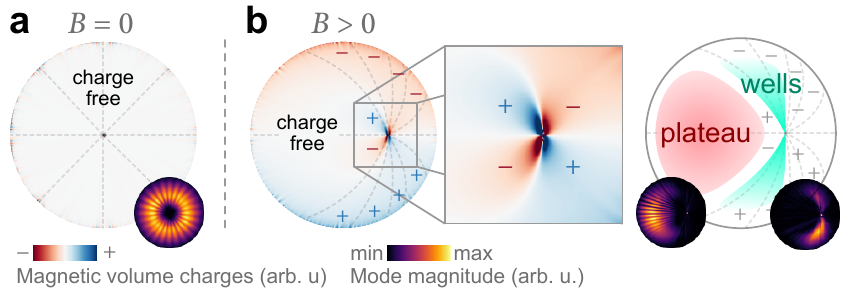}
    \caption{Static magnetic-volume charges in a (a) centered vortex at zero in-plane field and a (b) flexed vortex at non-zero in-plane field. Whereas at zero field, only one class of modes is present, at finite field, the flexing of the vortex skirt leads to plateaus and wells in the internal dipolar energy of the vortex, leading to two different classes of modes (\textit{regular} on the plateau and \textit{butterfly}-shaped in the wells).}
    \label{fig:FIG3}
\end{figure}

\begin{figure*}
    \centering
    \includegraphics{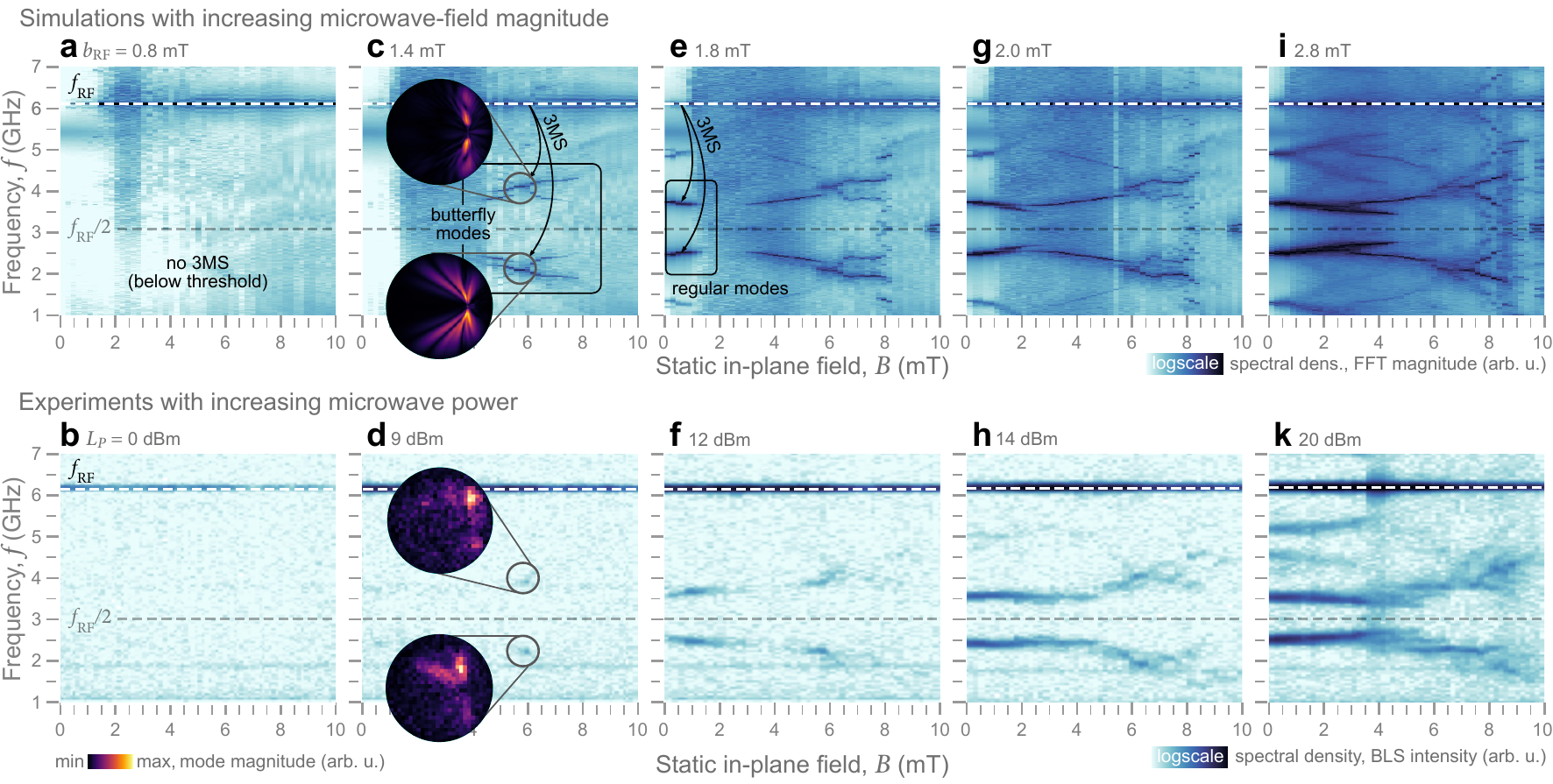}
    \caption{Evolution of the numerically and experimentally obtained frequency response of the magnetic vortex excited at a fixed microwave-excitation frequency of \SI{6.1}{\giga\hertz} with increasing magnitudes/powers of the driving microwave source for in-plane fields between 0 to \SI{10}{\milli\tesla}. For low excitation (a) and (b) only the primary magnon mode can be seen. At slightly higher excitations, panel (c) and (d) secondary modes appear for static fields for which the vortex is strongly distorted. The mode profiles obtained both from simulations and experiments are shown as insets and demonstrate that these modes are butterfly-shaped modes localized to the highly flexed regions of the vortex skirt. Further increasing the excitation strength, shown in  (e) and (f), leads to the appearance of the 3MS for regular vortex modes close the zero in-plane field, confirming that the three-magnon splitting into butterfly modes has a lower threshold power.  Finally, at large excitation strengths a rich and complex frequency response is measured, summarized on panels (i)-(k).
}
    \label{fig:FIG4}
\end{figure*}

The splitting of the spectrum of secondary modes into two classes can be understood in a simple picture by considering the landscape of magnetic volume charges created by deforming the magnetic vortex. To illustrate this, in Fig.~\ref{fig:FIG3}, we show the volume charges for a centered and a deformed vortex obtained from our simulations. At zero field [Fig.~\figref{fig:FIG3}{a}], the vortex achieves to suppress volume charges almost everywhere by forming a perfectly azimuthal rotation of in-plane magnetization, except close to the vortex core. As a result, the spectrum is only comprised of the well-known regular vortex modes described by azimuthal and radial mode numbers. For finite static in-plane fields, the flexing of the vortex skirt leads to an accumulation of magnetic volume charges adjacent to the core and at the boundaries of the magnetic disk close to the core, as seen in Fig.~\figref{fig:FIG3}{b}. As shown in Fig.~\figref{fig:FIG3}{c}, this charge distribution leads to potential wells for bound magnon state similar to the channeled magnons in Néel-type domain walls.\cite{wagnerMagneticDomainWalls2016} These bound modes have already been observed with low number of periods around the vortex core in Refs.~\citenum{alievSpinWavesCircular2009,jenkinsElectricalCharacterisationHigher2021}. In said works, the authors used linear excitation with in-plane microwave fields or direct spin-polarized currents to excite these modes. Both works, however, did not experimentally verify the spatial character of these modes. Apart from the flexed part of the vortex skirt, an (almost) charge-free region remains at the opposite side of the disk, leading to a potential plateau for the regular "free" vortex modes.

In the following we will see that the observed bound/butterfly modes exhibit a much lower three-magnon power threshold than the regular modes in a centered vortex. Again, we sweep the static in-plane field from 0 to \SI{10}{\milli\tesla}, however, this time, for different magnitudes/powers of the driving microwave field. We start with a microwave power below the nonlinear 3MS threshold, at $b_\mathrm{RF}=\SI{0.8}{\milli\tesla}$. We see in Fig.~\figref{fig:FIG4}{a,b} (simulation and experiment, respectively) that only the primary magnon at $f_\mathrm{RF}=\SI{6.1}{\giga\hertz}$ is excited over the whole field range and no additional secondary modes. With increasing microwave power [Fig.~\figref{fig:FIG4}{c,d}], secondary modes appear for static fields $\gtrsim\SI{4.5}{\milli\tesla}$ at which the vortex is deformed considerably. Analyzing the spatial mode profiles of these modes from experiments and simulation reveals that all of them are indeed modes localized to the highly flexed regions of the vortex skirt [shown as insets in Fig.~\figref{fig:FIG4}{c,d}]. Note again, that all of them are arranged in pairs with frequency $f_{1,2}=f_\mathrm{RF}/2 \pm \Delta f$ and different "radial" profiles. Only when increasing the microwave power even further, the 3MS threshold of the regular vortex modes is reached [Fig.~\figref{fig:FIG4}{e,f}], and their frequency contribution is visible at small in-plane fields at which the vortex is not deformed at all. Finally, with further increasing the microwave power, the branches of the two classes of modes approach each other until their overlap in in-plane-field range and ultimately, the behavior discussed in context of Fig.~\ref{fig:FIG2} is recovered.

The reason of the butterfly modes to inherit a lower three-magnon threshold, compared to the regular vortex modes, can be two-fold: First, their intrinsic damping rates $\Gamma_{1,2}$~\cite{} could be significantly lower which would point towards lower mode ellipticities, or, second, their three-magnon scattering coefficients $V_{0,12}$~\cite{} with respect to the directly-excited (primary) magnon are significantly larger.\cite{verbaTheoryThreemagnonInteraction2021} A combination of both factors finally results in lower threshold microwave fields $b_{\mathrm{RF},\mathrm{crit}}\propto \Gamma_1\Gamma_2/\abs{V_{0,12}}$. To answer this open question, an analysis of both factors separately would be necessary, which would be possible, \textit{e.g.} within the frame of a vector Hamiltonian formalism for nonlinear magnon dynamics\cite{tyberkevychVectorHamiltonianFormalism2020} after having numerically calculated the spatial mode profiles, \textit{e.g.} with a dynamic-matrix method.\cite{grimsditchMagneticNormalModes2004} This would, however, go beyond the scope of this work.

With our study of three-magnon splitting in a flexed magnetic vortex, confined to a micrometer-sized magnetic disk,  we have shown that this nonlinear process is stable with respect to displacement of the vortex core by in-plane magnetic bias fields. Application of such fields leads to a flexing of the vortex skirt, which, in return, leads to a separation of the secondary modes produced by three-magnon splitting into modes corresponding to the regular radial and azimuthal modes of a symmetric vortex, and, into additional \textit{butterfly} modes which are confined to the highly flexed regions adjacent to the vortex core. Splitting into these additional secondary modes exhibits a much lower power threshold than into the regular vortex modes. This work expands the understanding of three-magnon splitting in confined magnetic systems, providing a way to excite unconventional magnon modes in flexed magnetic vortices. Furthermore, small in-plane bias fields, in the range of a few tens of \si{\milli\tesla} are shown to be a powerful parameter to tune the characteristics of a microscopic nonlinear systems in place, which is attractive, for example, for neuromorphic applications such as reservoir computing.

\section*{Author declarations}

\subsection*{Conflict of Interest}
The authors have no conflicts of interest to disclose.

\subsection*{Author's contributions}

\textit{Will be inserted via submission form.}

\section*{Acknowledgements}

The authors acknowledge fruitful discussions with V. Tyberkevych. Financial support by the Deutsche Forschungsgemeinschaft (DFG) within the programs SCHU 2922/1-1, KA 5069/1-1 and KA 5069/3-1 is gratefully acknowledged as well as from the EU Research and Innovation Programme Horizon Europe under grant agreement no. 101070290 (NIMFEIA). Support by the Nanofabrication Facilities Rossendorf (NanoFaRo) at the IBC is gratefully acknowledged.

\section*{Data availability}
The data that support the findings of this study are openly available in RODARE.\footnote{L. Körber, C. Heins, I. Soldatov, R. Schäfer, A. Kákay, H. Schultheiss, K. Schultheiss, `` Data publication: Modification of three-magnon splitting in a flexed magnetic vortex,``  RODARE \url{http://doi.org/10.14278/rodare.2064}, version 1 2022}



%

\end{document}